\documentclass[aip,apl,preprint,amsmath,amssymb]{revtex4-1}

\usepackage{amsmath}
\usepackage[utf8]{inputenc}
\usepackage{wasysym}
\usepackage{upgreek} 
\newcommand{\degre}{^\circ}
\newcommand{\cminv}{\, \mathrm{cm}^{-1}}
\newcommand{\microns}{\, \upmu \mathrm{m}}

\newcommand{\mw}{\, \mathrm{mW}}
\newcommand{\mm}{\, \mathrm{mm}}
\newcommand{\ma}{\, \mathrm{mA}}

\usepackage{graphicx}
\begin{document} 



\title{Watt-level widely tunable single-mode emission by injection-locking of a multimode Fabry-Perot quantum cascade laser}



\author{Paul Chevalier}
\affiliation{Harvard John A. Paulson School of Engineering and Applied Sciences, Harvard University, Cambridge, MA 02138 USA}

\author{Marco Piccardo}
\affiliation{Harvard John A. Paulson School of Engineering and Applied Sciences, Harvard University, Cambridge, MA 02138 USA}

\author{Sajant Anand}
\affiliation{Harvard John A. Paulson School of Engineering and Applied Sciences, Harvard University, Cambridge, MA 02138 USA}
\affiliation{Department of Physics, Wake Forest University, Winston-Salem, NC 27109, USA}

\author{Enrique A. Mejia}
\affiliation{Harvard John A. Paulson School of Engineering and Applied Sciences, Harvard University, Cambridge, MA 02138 USA}
\affiliation{University of Texas at Austin, Austin, TX 78712, USA}

\author{Yongrui Wang}
\affiliation{Department of Physics and Astronomy, Texas A\&M University, College Station, TX 77843 USA}

\author{Tobias S. Mansuripur}
\affiliation{Pendar Technologies, 30 Spinelli Pl, Cambridge, MA 02138 USA}

\author{Feng Xie}
\affiliation{Thorlabs Quantum Electronics (TQE), Jessup, MD 20794 USA}

\author{Kevin Lascola}
\affiliation{Thorlabs Quantum Electronics (TQE), Jessup, MD 20794 USA}

\author{Alexey Belyanin}
\affiliation{Department of Physics and Astronomy, Texas A\&M University, College Station, TX 77843, USA}

\author{Federico Capasso}
\email[]{capasso@seas.harvard.edu}
\affiliation{Harvard John A. Paulson School of Engineering and Applied Sciences, Harvard University, Cambridge, MA 02138 USA}


\date{\today}

\begin{abstract}
 Free-running Fabry-Perot lasers normally operate in a single-mode regime until the pumping current is increased beyond the single-mode instability threshold, above which they evolve into a multimode state. As a result of this instability, the single-mode operation of these lasers is typically constrained to few percents of their output power range, this being an undesired limitation in spectroscopy applications. In order to expand the span of single-mode operation, we use an optical injection seed generated by an external-cavity single-mode laser source to force the Fabry-Perot quantum cascade laser into a single-mode state in the high current range, where it would otherwise operate in a multimode regime. Utilizing this approach we achieve single-mode emission at room temperature with a tuning range of $36 \cminv$ and stable continuous-wave output power exceeding 1 W at $4.5 \microns$. Far-field measurements show that a single transverse mode is emitted up to the highest optical power indicating that the beam properties of the seeded Fabry-Perot laser remain unchanged as compared to free-running operation.
\end{abstract}


\maketitle
Several spectroscopy applications, such as remote chemical sensing of atmospheric gases~\cite{xie2012room}, photoacoustic spectroscopy~\cite{ma2013qepas} and real-time infrared imaging of living micro-organisms~\cite{haase2016real}, require short acquisition times or long-distance propagation, setting a demand for continuous wave (CW), high-power, single-mode lasers. 
Quantum cascade lasers~\cite{faist1994quantum} (QCLs) have reached in the past years record CW power output\cite{figueiredo2017progress} with Fabry-Perot (FP) devices~\cite{yao2010high,lyakh2012tapered,razeghi2014recent} emitting more than 5 W, and distributed-feedback (DFB) devices~\cite{lu20112} emitting up to 2.4 W. FP lasers have the simplest cavity geometry consisting of only a waveguide and two reflective interfaces; however, since the cavity supports multimode operation, a FP-QCL does not remain single-mode over more than typically a few percent of its power output range~\cite{mansuripur2016single}.
On the other hand DFB lasers include gratings in their waveguide in order to support only single-mode operation; however, this comes at the cost of a more complex fabrication process, and their power output is usually lower than their FP counterparts.
Careful engineering of the DFB grating also has to be done to ensure a proper single-lobe beam~\cite{lu20112}.
More recently, double-section devices consisting of a DFB QCL section coupled to an amplifier section~\cite{rauter2015multi,bismuto2016high} have been shown to emit up to 1 W single-mode under the appropriate driving conditions. However, tuning of the emission wavelength of DFB-based devices relies on controlling their temperature and results in a strong variation of the output power as a function of the tuned wavelength.
In laser diodes, injection-locking has been used to control the spectral properties\cite{kobayashi1980injection,goldberg1987injection,kim2000low} of the device. This technique consists of injecting light from a master source into a slave laser to lock the spectral emission of the latter. Injection-locking has also been demonstrated in QCLs with mutually coupled Fabry-Perot and DFB lasers, allowing to achieve a total output power of the order of $40 \mw$~\cite{bogris2017mid}, or to improve the stability of a QCL~\cite{juretzka2015intensity,taubman2004stabilization}.

In this letter, we propose a different approach than DFB lasers, and obtain high-power single-mode emission from a FP-QCL by injecting an optical seed. We first discuss the limits of the single-mode regime of free-running FP-QCLs in relation to their facet coatings. Then by using an external cavity QCL, we demonstrate injection-locked single-mode operation of a FP laser, resulting in up to 1 W of optical power emitted by the laser at $4.5 \microns$, with a tuning range of $36 \cminv$. The power of the optical seed coupled into the waveguide of the FP laser does not exceed few mW. An insight into the evolution from multimode emission to single-mode operation is provided by time-resolved space-time domain QCL simulations~\cite{Wang:15}. Finally, far-field measurements are presented showing that the injected laser exhibits a single-lobe beam, indicating that the beam quality is not modified by this injection scheme.

The typical spectral evolution observed in CW FP lasers is the following: just above threshold, the laser operates in single-mode; as the pumping current is further increased, the emission spectrum evolves into a multimode state (either a dense state with adjacent cavity modes populated or a harmonic state with modes separated by several cavity free spectral ranges). The evolution into a multimode state results from the single-mode instability and is explained by a combined effect of spatial hole burning and population pulsation nonlinearity~\cite{mansuripur2016single}.
\begin{figure}[t]
\centering
\includegraphics[width=0.37\textwidth]{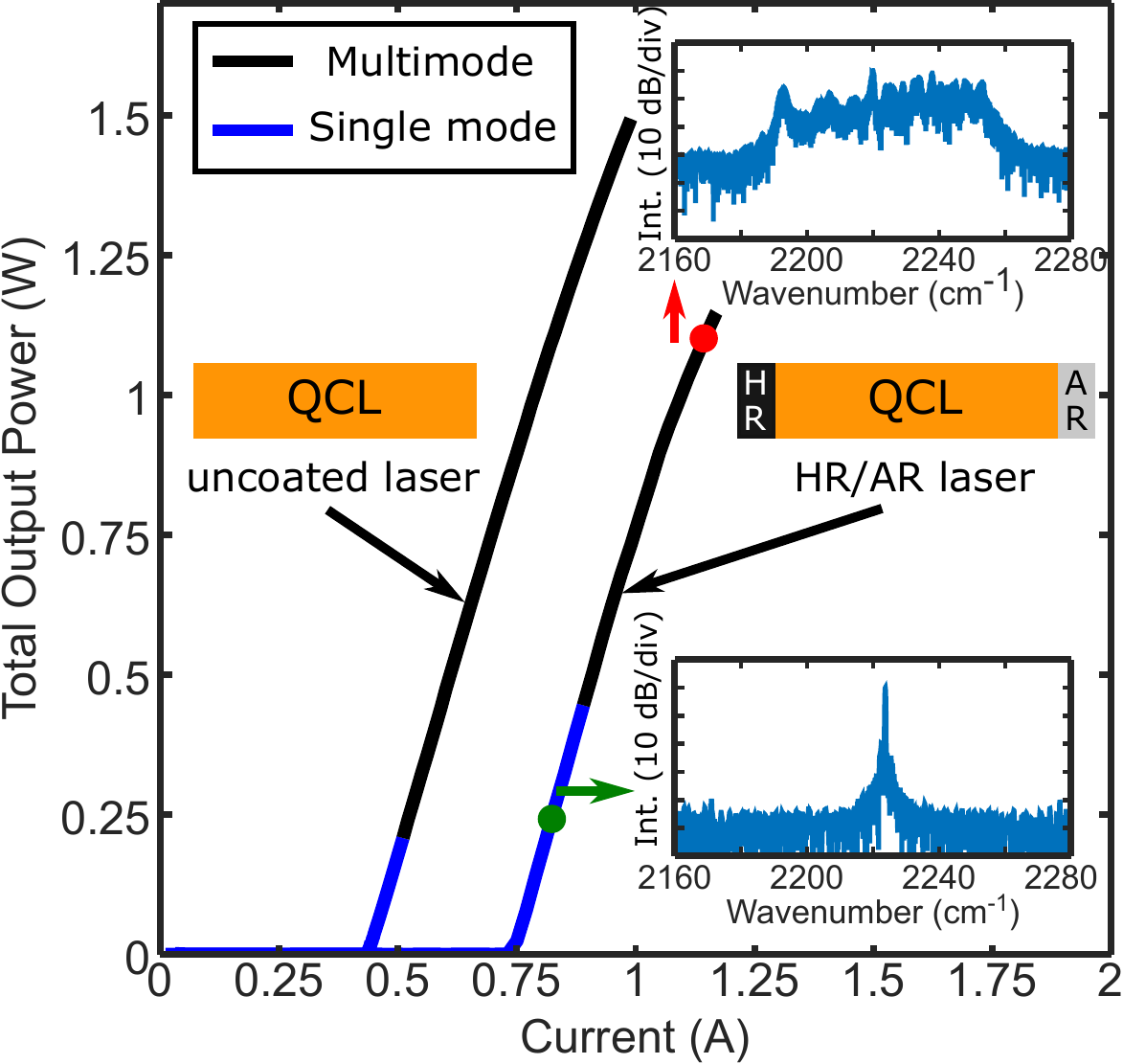}
  \caption{\label{fig:fig1}
Laser output power as a function of the driving current showing the range of each emission state of the laser (single-mode or multimode). The plot compares the power curves of an uncoated laser and a HR/AR-coated laser, with identical active regions and ridge geometries.
Inset: Output spectrum for the HR/AR laser plotted at two driving currents corresponding to the single-mode and multimode dense regime ($850 \ma$ and $1200 \ma$, respectively).
}
\end{figure}
In Fig.~1 we plot the total output power curve of two FP lasers (cavity length: $6 \, \mathrm{mm}$, width: $5 \microns$) fabricated from the same process and which differ only in the facet coatings. One laser has uncoated facets while the other has anti-reflective (AR, $R \approx 1\%$) and highly-reflective (HR, $R \approx 99\%$) coatings on each respective facet. In this figure we emphasize the dynamic range of the single-mode regime (red part of the curve) and that of the multimode regime (gray part of the curve). In the inset of Fig. 1, the spectra of the HR/AR coated laser in single-mode (pumping current: $850 \ma$) and multimode (pumping current: $ 1200 \ma$) regime are shown.
The power-current curves also show that the laser with HR/AR coatings remains single-mode over a broader range of emitted power (up to $450 \mw$) than the uncoated laser (up to approximately $100 \mw$ per facet). This behavior was explained in the theoretical study reported in Ref.~\citenum{mansuripur2016single}: the HR/AR coatings transform the standing-wave cavity into more of a traveling-wave cavity, thus the single-mode instability occurs at a higher current. 

\begin{figure*}[htp]
\centering
\includegraphics[width=0.8\textwidth]{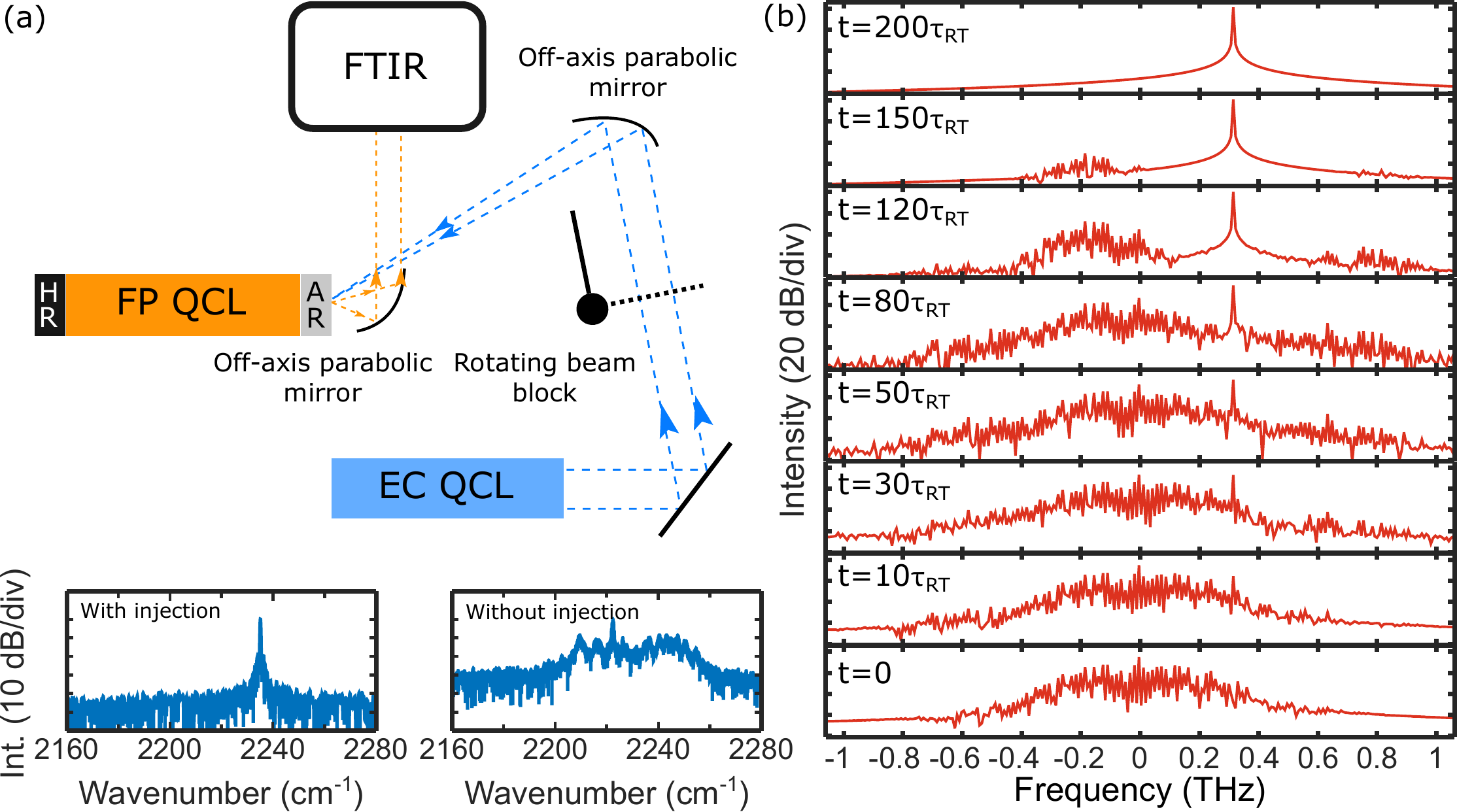}
  \caption{\label{fig:fig2}
(a) Schematic of the injection-locking setup: light from a tunable external-cavity (EC) QCL is injected at an angle in a HR/AR coated Fabry-Perot (FP) QCL device with an off-axis parabolic mirror. The spectrum of the slave laser is measured with a Fourier-transform infrared (FTIR) spectrometer. The beam of the EC-QCL can be blocked with a beam stopper mounted on a computer-controlled rotational stage.
The insets show the emission spectra of the FP laser in two cases: (left) light from the EC-QCL is unblocked and the FP laser spectrum is single-mode, (right) light from the EC-QCL is blocked and the FP laser spectrum is multimode.
(b) Simulated time-resolved evolution of the emission spectrum from a FP-QCL after optical injection with a single-mode detuned by $0.32 \, \mathrm{THz}$ from the gain peak. The x-axis corresponds to the frequency with respect to the FP-QCL gain peak position. Each spectrum is calculated from the emitted electric field over a 2$\tau_{\mathrm{RT}}$-long time window, where $\tau_{\mathrm{RT}}$ is the cavity roundtrip time. An animated version of the time-resolved spectral evolution can be viewed in the multimedia video provided in the online version of the paper. (Multimedia view).
}
\end{figure*}

The previous characterization shows that the free-running FP laser does not remain single-mode for the complete extent of its power range due to the occurrence of the single-mode instability. In order to broaden the power range of single-mode operation, our approach consists of injecting a light seed generated by a tunable single-mode laser into the cavity of the FP-QCL to prevent its operation in a multimode regime.
The schematic of the experimental setup is shown in Fig.~2(a). The FP device is the HR/AR laser presented in Fig.~1. Its emitted light is collimated using an off-axis parabolic mirror (diameter $12.7 \mm$, focal length $15 \mm$) and sent into a Fourier transform infrared spectrometer (FTIR, Bruker 70). Light from a tunable external cavity (EC) QCL is coupled into the FP-QCL at a $34^{\circ}$ angle by another off-axis parabolic mirror (diameter $50 \mm$, focal length $50 \mm$). The EC-QCL is a commercial laser (41045-HHG, Daylight Solutions) whose single-mode output can be tuned between $2170 \cminv$ and $2320 \cminv$ with a maximum output power of $350 \mw$. The coupling efficiency of the injection setup was estimated using the FP-QCL as a mid-infrared detector and comparing the voltage across it resulting from the injection of EC-QCL light at normal and oblique incidence. Assuming that the relative change in QCL responsivity among the two configurations is equal to the relative change in coupling efficiency, and assuming a unity coupling efficiency at normal incidence (a generous estimate considering that the focusing is not diffraction-limited, giving a focal spot larger than the waveguide size), we deduce an upper bound for the coupling efficiency at a $34^{\circ}$ angle of $1.25 \%$, resulting in at most $4.4 \mw$ of optical power coupled into the FP device. This experimental arrangement, in which light is injected at an angle, despite its low coupling efficiency, was chosen to reduce the amount of light sent into EC-QCL by the FP-QCL, thus providing an effective optical isolation scheme. Additionally a beam stopper mounted on a computer-controlled stage was added on the beam path of the EC-QCL to allow or block injection of light into the FP-QCL. The injected beam was blocked every time the emission wavelength and power of the EC QCL was tuned or when the current of the FP laser was changed. 

The EC-QCL was tuned to $2235 \cminv$ and the pump current of the FP laser was progressively increased above $1 \, \mathrm{A}$. At each current step the light from the EC-QCL was unblocked, then the emission spectrum of the FP-QCL was measured and finally the light was blocked again. When the FP-QCL current reached $1050 \ma$ we measured a single-mode spectrum as shown in the inset of Fig.~2(a). The output power of the FP-QCL at this pumping current was slightly above 1~W. To assess the stability of such a single-mode state, the beam from the EC-QCL was successively blocked and unblocked multiple times while the emission spectrum of the FP-QCL was measured. The FP-QCL always returned to a multimode state (spectrum shown in the inset of Fig.~2(a)) after the seed was blocked. The emission spectrum of the FP-QCL became single-mode again after the beam was unblocked, thus showing the repeatability and reversibility of the process. We also studied the influence of the injected power on single-mode operation while maintaining the FP-QCL current at $1050 \ma$.  The injected power was adjusted between $0.36 \mw$ and $3.4 \mw$: below $0.46 \mw$ only multimode operation of the FP-QCL was observed. Above this critical power we were able to lock the laser into a single-mode, and as the injected power was increased up to $3.4 \mw$, the measured output intensity did not increase.

In order to understand the underlying physics of the observed single-mode behavior, the transition from multimode to single-mode in the FP-QCL upon optical injection should be analyzed. While such measurement is experimentally challenging, the transient regime can be studied by space- and time-domain simulations of the FP-QCL in presence of an optical seed. The model used is described in Ref.~\citenum{Wang:15} and the characteristic parameters are: the coupling energy between the injector state and the upper laser state: $\mathrm{\Omega=2.25\,meV}$, the lifetime from the upper laser state to the lower laser state: $\mathrm{T_{ul}=1\,ps}$, the lifetime from the upper laser state to the ground state: $\mathrm{T_{ug}=3\,ps}$, the lifetime from the lower laser state to the ground state: $\mathrm{T_{lg}=0.1\,ps}$, the dephasing time for the polarization between upper and lower laser states: $\mathrm{T_2=0.05\,ps}$,  the diffusion coefficient for electrons in each state $\mathrm{D}=46 \, \mathrm{cm^2/s}$, the cavity losses: $\mathrm{l_w=5\,cm^{-1}}$, the mode overlap: $\mathrm{\Gamma=0.5}$, the facet reflection coefficients: $\mathrm{R_l=1}$ and $\mathrm{R_r=0.1}$. The other parameters of the model are the same as in Ref.~\citenum{Wang:15} and a schematic of the energy levels of the active region can be seen in Fig.~1(a) therein. The seed has an amplitude of 1.0 kV/cm, a frequency detuning of 0.32\,THz, and is turned on at $\mathrm{t}=0$. Before that, the laser operates in a dense multimode state (see Fig.~2(b)). The laser evolves into single-mode state in about 200 roundtrips, as shown in Fig. 2(b) (Multimedia view). Based on the simulation, we explain the dynamics of the optically-injected FP-QCL as follows. As the power of the seed is chosen to be slightly larger than that of the lasing modes reflected from the AR facet, the seed, which propagates as a running wave, competes with other longitudinal modes. This leads to the depletion of the gain of other longitudinal modes (see, for instance, Fig.~2(b), $t=120\tau_{\mathrm{RT}}$) and suppresses them, so that a single-mode regime may be established after many round-trips. The scheme presented here is similar to injection-locking as described in Ref.\citenum{siegman1986lasers}, however, when compared to the text-book case of injection-locking, our FP-QCL emits a multimode state when it is not optically injected.

\begin{figure}[htp]
\centering
\includegraphics[width=0.38\textwidth]{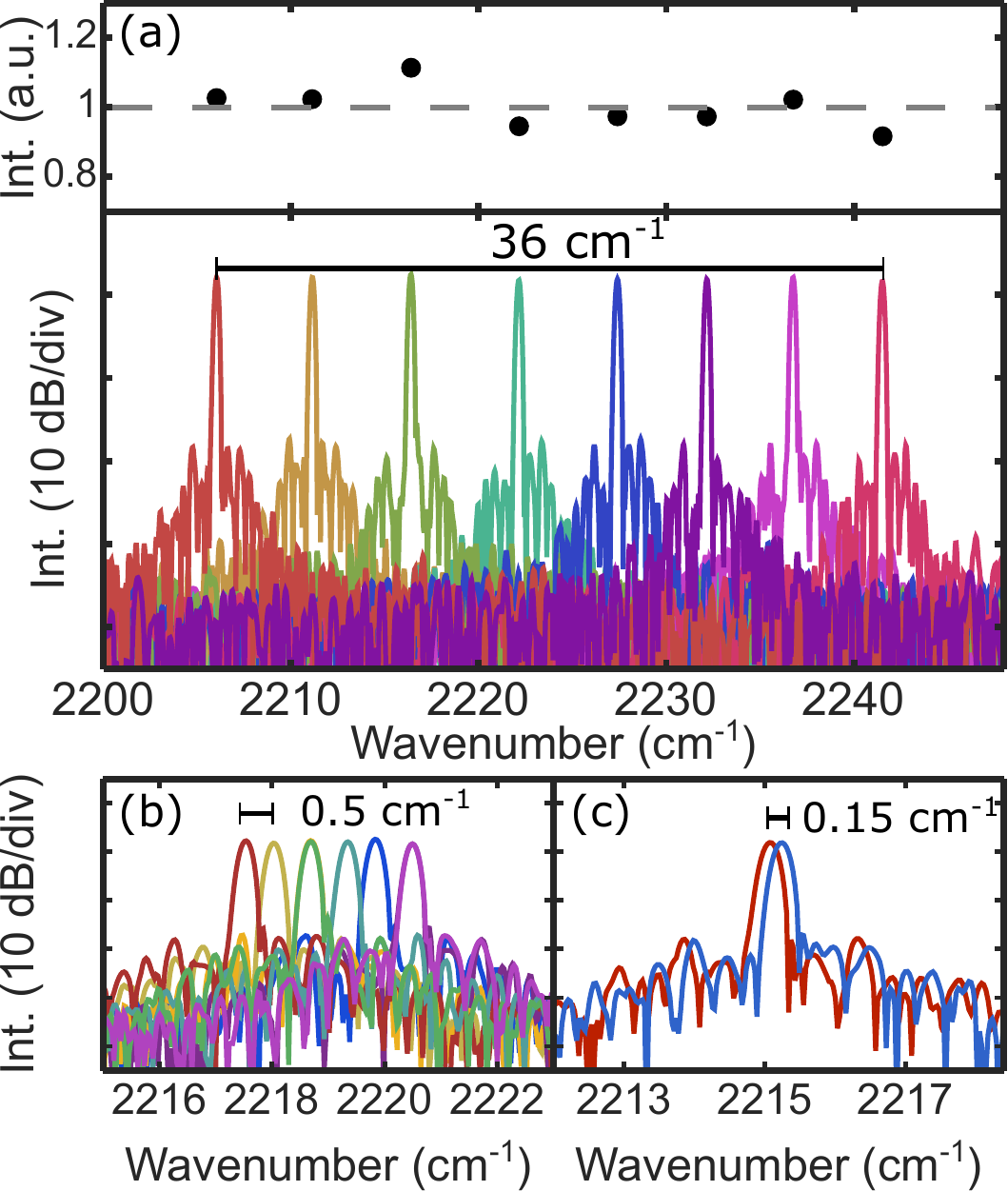}
  \caption{\label{fig:fig3}
Measured single-mode emission spectra from the FP-laser for different tuning ranges of the EC-QCL:
(a) Bottom: Tuning the emission of the FP laser over a range $36 \cminv$. Top: Intensity of the single-mode emission plotted on a linear scale as a function of the emission wavenumber showing a variation of less than 20\% over $36 \cminv$.
(b) Tuning the emission of the FP laser by steps of  $0.5 \cminv$.
(c) Tuning the emission of the FP laser by $0.15 \cminv$.
}
\end{figure}

The emission of the FP-QCL can be experimentally tuned over a range of $36\cminv$ using the following protocol. First the EC-QCL has to be set to the desired emission frequency. 
Then the pumping current of the FP-QCL is adjusted in order to thermally-tune the position of the cavity modes, so that the injected seed corresponds to one of them. 
By applying this tuning protocol, we demonstrate in Fig. 3(a-c) the tuning range of the injection-locked single-mode operation. Even if the injection-locked scheme allows to tune the emission of the FP-laser close to one of its mode frequencies, the latter can be tuned by adjusting the pumping current, thus allowing any arbitrary single-mode frequency to be emitted without significant variation of the output power.

In Fig.~3(a) the output spectra of the FP-QCL are shown for various emission wavenumbers of the EC-QCL: this plot shows that upon injection of the single-mode from the EC-QCL, the output spectrum of the FP-QCL can be tuned over $36 \cminv$ while remaining single-mode. The top part of Fig.~3(a) shows the intensity single-mode emission on a linear scale as a function of the emitted wavenumber. 
We show in Fig.~3(b) (respectively in Fig.~3(c)) that the emission wavelength of the FP-QCL can be tuned with smaller steps: five different modes separated by $0.5 \cminv$ (respectively two modes separated by $0.15 \cminv$) were produced using different driving conditions of the lasers. For all the tuning experiments, the value of the FP-QCL current was set between $1050 \ma$ and $1100 \ma$ in order to safely operate the laser well below its roll-over, still these values were sufficient to demonstrate a single-mode operation with a 1~W output power.

\begin{figure}[htp]
\centering
\includegraphics[width=0.37\textwidth]{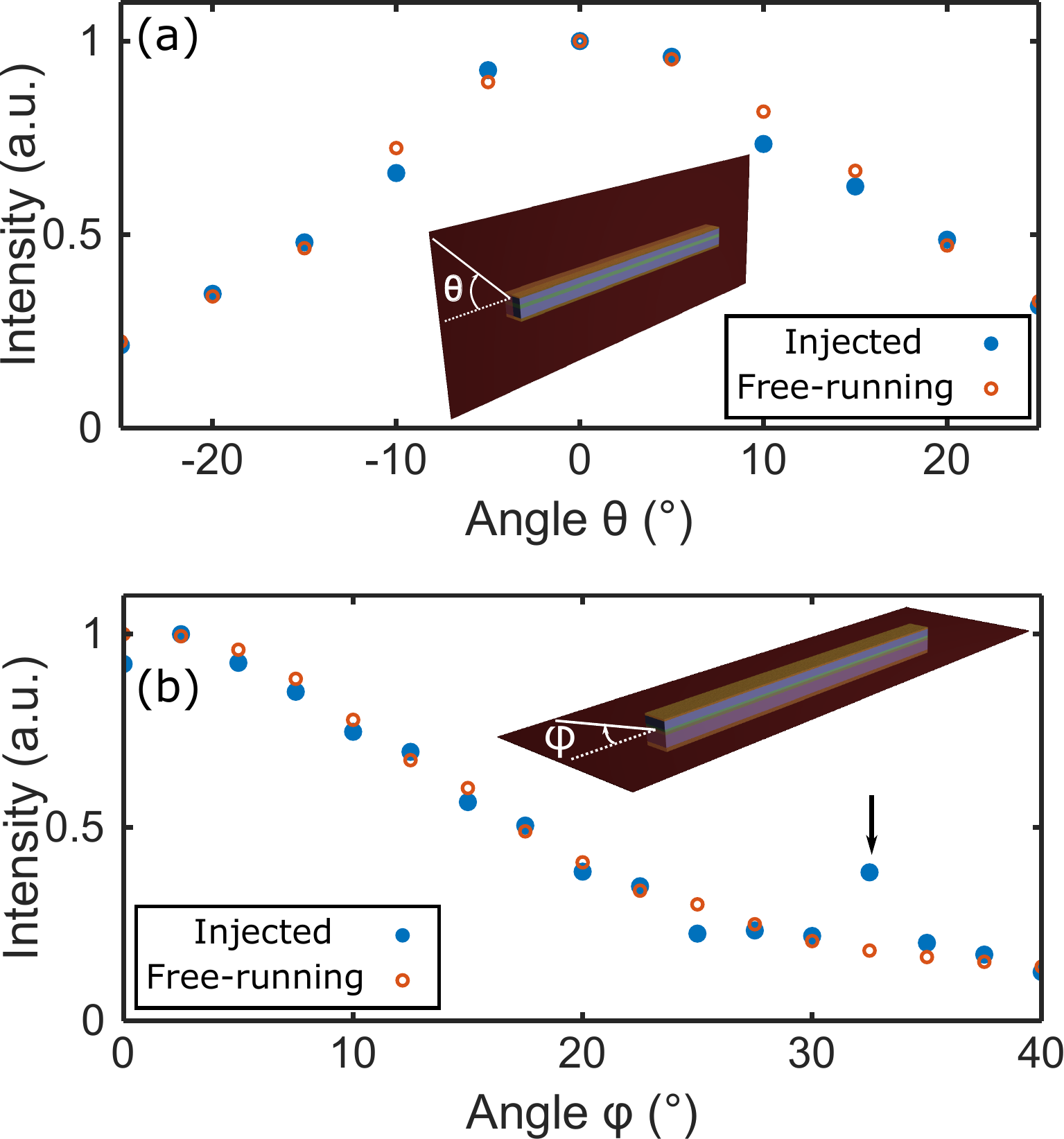}
  \caption{\label{fig:fig4}
Measurement of the far-field radiation of the laser in the single-mode state, and in free-running mode at the same driving current ($1050 \ma$):
(a) far field measured in the transverse plane;
(b) far field measured in the longitudinal plane. The arrow indicates the peak of intensity due to the reflected seed on the facet of the FP-QCL.
}
\end{figure}

Since the EC-QCL light is injected at an angle by the off-axis parabolic mirror, we wanted to determine whether this scheme causes the FP-laser to emit light into higher-order transverse modes, thus degrading the beam quality. To characterize the beam, we then measured the far-field emission of the FP-QCL in both free-running and injection-locked single-mode regime. The FP-QCL was driven by a current of $1085 \ma$ and injection-locked at $2235 \cminv$.  The collimating parabolic mirror (see Fig.~2(a)) was removed and the far-field radiation was measured in the transverse plane (blue curve in Fig.~4(a)) and in the longitudinal plane (blue curve in Fig.~4(b)) by using a thermoelectrically-cooled HgCdTe photodetector (Vigo PVI-4TE). The same measurements were repeated with the EC-QCL turned-off to measure the far-field of the FP-QCL without injection (red curves in Fig. 4). The two far-field profiles do not show significant changes except for a peak (see arrow in Fig. 4(b)) in the longitudinal plane around $34 \degre$ corresponding to the light from the EC-QCL reflected by FP-QCL facet. This measurement shows that even if the injection is at oblique incidence, the single-mode operation induced by the technique reported here corresponds to the fundamental transverse mode (TM$_{00}$) of the FP-QCL waveguide. 

Compared to other integrated systems using a DFB master oscillator and generating similar or higher power\cite{lu20112,bismuto2016high}, our technique ensures that the maximum power of the FP laser is extracted over the whole emission range ($36 \cminv$) into single-mode emission, whereas the tuning of a DFB master oscillator only achieves a narrower tuning thanks to a significant variation of the temperature and thus results in uneven power over the emission range.

To conclude, we demonstrated here a method to achieve, through optical injection, powerful and widely-tunable single-mode operation of a FP-QCL, a system that would otherwise emit a multimode spectrum. The single-mode operation was achieved with a total output power of 1~W by injecting less than 1~mW of light into the slave laser and was demonstrated over a range of $36 \cminv$ (between $2206 \cminv$ and $2242 \cminv$). Insight into the evolution of the laser upon optical injection was provided by time-resolved space-time domain QCL simulations. The proposed scheme is not necessarily limited to a tabletop setup: one may envision realizing a compact source by using an array of DFB lasers all injected into a single FP-QCL, where the emission wavelength of the DFBs can be adapted to the targeted spectral range.  In such configuration the master and the slave lasers will work in tandem, since the FP-QCL would provide high optical power and coupling through a single device and the DFB array the necessary wavelengths.
\begin{acknowledgments}
This work was supported by the DARPA SCOUT program through Grant No. W31P4Q-16-1-0002. We acknowledge support from the National Science Foundation under Award No. ECCS-1614631. Any opinions, findings, conclusions or recommendations expressed in this material are those of the authors and do not necessarily reflect the views of the Assistant Secretary of Defense for Research and Engineering or of the National Science Foundation. We thank D. Kazakov for a careful reading of this manuscript.
\end{acknowledgments}

\end{document}